# Determination of the Decay Exponent in Mechanically Stirred Isotropic Turbulence

## Blair Perot

**Abstract**
Direct numerical simulation is used to investigate the decay exponent of isotropic homogeneous turbulence over a range of Reynolds numbers sufficient to display both high and low Re number decay behavior. The initial turbulence is generated by the stirring action of the flow past many small randomly placed cubes. Stirring occurs at $1/30^{th}$ of the simulation domain size so that the low-wavenumber and large scale behavior of the turbulent spectrum is generated by the fluid and is not imposed. It is shown that the decay exponent in the resulting turbulence closely matches the theoretical predictions of Saffman (1967) at both high and low Reynolds numbers. The transition from high Reynolds number behavior to low Reynolds number behavior occurs relatively abruptly at a turbulent Reynolds number of around 250 ($\text{Re}_\lambda \approx 41$).

## 1. Introduction

Decaying isotropic turbulence is perhaps the simplest and most analyzed turbulent flow. Nevertheless, there remains considerable debate about one of its most basic properties - the decay rate. This debate might be of academic consequence if it were not for the fact that lack of decay rate certainty hinders the development of turbulence models. Decay constants are typically set first in a turbulence model and incorrect or imprecise specification of the decay rate then results in a cascade of compensating errors for all the other model terms. This paper therefore seeks to obtain more certainty about the decay rate in isotropic turbulence.

A running debate about the isotropic turbulent decay rate remains in force primarily for two reasons. First, the decay rate is an extremely sensitive parameter to measure. It is arguable that many prior experiments lack sufficient resolution and/or sufficient duration to make sufficiently precise statements about decay rates to resolve the debate. Secondly, classical numerical experiments or direct numerical simulations (DNS), which can obtain the necessary resolution and duration, have decay rates which are dictated by the input initial conditions. Neither experiments nor prior numerical simulations have therefore adequately resolved the disagreement among the differing theoretical predictions.

In this work, turbulence is generated (and not imposed as an initial condition) by moving initially zero velocity fluid past a collection of randomly placed objects (small cubes). The turbulence being studied in these simulations is therefore only a result of the Navier-Stokes equations. The turbulence is not a result of external forcing or the initial conditions (which are zero). It is only the result of 'stirring' by small rigid objects. It was decided not to simulate the wind tunnel generation mechanism (flow past a rigid grid of wires) such as was performed by Djenidi (2006) because this does not create isotropic turbulence. Both the fluctuations and the structure of wind tunnel grid turbulence are axisymmetric around the flow direction. The use of a contraction to remove the fluctuation anisotropy (in the Reynolds stresses) does not remove the structural anisotropy in this flow configuration. Therefore the two-point correlations, on which all the decay theories are predicated, remain anisotropic (Kurian and Fransson 2009). Because the decay exponent is an extremely sensitive quantity, the wind tunnel configuration (and its anisotropy) therefore adds an addition level of uncertainty that would undermine the results. With high resolution direct numerical simulation of isotropic turbulence it is hoped that we can identify which of the many theories concerning isotropic decay is applicable to turbulence produced by mechanical stirring.  Note that there are many turbulence generation mechanisms (buoyancy, gravity waves, etc) and they might well have different decay behaviors.  The focus of this work is on the decay exponent of turbulence that is generated by the presence of solid objects.

**1.1 Decay Theories**
All theories concerning isotropic decay (von Karman & Howarth 1938, Kolmogorov 1941, Saffman 1967, George 1992, Speziale & Bernard 1992) agree that the kinetic energy $K$, should have a power law decay behavior in time, $t$,

$$K(t) = \hat{K}(t+t_0)^{-n} \qquad (1)$$

where the magnitude, $\hat{K}$, the time offset, $t_0$, and the decay exponent, $n$, are constants. The dimensional parameters $\hat{K}$ and $t_0$ are expected to depend on the particular flow situation, but the dimensionless exponent $n$ is expected to be universal for all isotropic decaying flows.  The disagreement between theories is therefore over the value of the decay exponent *n*.  There are multiple theories for decay exponent's value in both the high Reynolds number asymptotic limit and in the low Reynolds number asymptotic limit which is sometimes called 'the final period of decay'.

Batchelor and Townsend (1948b) presented one of the first analyses and experiments for the low Reynolds number turbulent decay regime.  This should be a limit in which very rigorous statements about turbulence ought to be able to be made because in theory the nonlinear terms become negligible and the equations become linear and solvable.  Indeed, Batchelor and Townsend derive an exact solution for the two-point correlation equation in the limit when the nonlinear terms are neglected. The decay rate corresponding to that exact solution is *n = 5/2*.  To confirm the theoretical analysis they performed experiments that indeed matched this 'final period' decay rate.  The high quality of this theoretical and experimental work means that this result (*n = 5/2*) remains pervasive in turbulence models and texts seven decades later.  The low Reynolds number value of *n=5/2* corresponds to a K/ε model decay constant of $C_{\varepsilon 2} = 1 + \frac{1}{n} = \frac{7}{5}$ which is found as the low Reynolds number limit in a large number of low Reynolds number turbulence model implementations (Launder and Sharma 1974, Lam and Bremhorst 1981, Chung & Kim 1998, and others).  Saffman (1967) later proposed that *n = 3/2* is also a possible solution at low Reynolds numbers.

The earliest high Reynolds number decay experiments (such as those by Batchelor and Taylor 1948a) hypothesized a high Reynolds number decay exponent of *n=1*.  Later it was shown by George (1992) and Speziale and Bernard (1992) that fully self-similar solutions to the full two-point evolution equation would require that *n=1*.  Lie group analysis of the Navier-Stokes equations (Oberlack, 2002) of course give the same result.  The possibility of *n=1* decay has even been recently reinvestigated with simulations by Burattini *et al.* (2006).  While the math itself is flawless, the basic assumption of self-similarity may well be flawed.  Self-similarity occurs in situations (such as a laminar jet) where no external length scale is active (except the one achieved by the solution).  High Reynolds number turbulence is a situation where at least two very distinct length scales can readily be identified (the integral scale and the viscous scale).  These scales correspond roughly to the peak in the spectrum and the spectrum cut-off.  In *n=1* decay the Reynolds number should be constant.  All experiments and simulations to date have shown a decaying Reynolds number with time, and so a decay exponent of *n=1* is not favored by most modelers.

Kolmorgorov (1941) determined a high Reynolds number theoretical value of *n=10/7* for the decay exponent based on the invariance of the Loitsyansky invariant.  Simulation data (Chasnov 1993) and EDQNM results (Lesieur 1987) suggest the Loitsyansky integral is not exactly an invariant and varies very slowly in time so the actual value is not 10/7 but 3% (Lesieur) to 6% (Chasnov) lower (1.38-1.34 instead of 1.428). Davidson (2004) and Ishida *et al.* (2006) have also carefully analyzed this situation, with similar conclusions. The principal contender to the Kolmogorov theory for high Reynolds number turbulence is an analysis by Saffman (1967) which at high Reynolds numbers determines the decay exponent to be *n=6/5*. If Saffman's theory is applicable, the Loitsyansky integral goes to infinity. Oberlack (2002) showed that the Saffman value of n=6/5 emerges from a Lie group analysis of the *Euler* equations (i.e. neglecting the influence of viscosity).

Finally, the possibility of *n=2* decay exists. At long times the decay rates are influenced by the simulation domain or wind tunnel cross section size. In decaying turbulence the integral scale (or large eddy scale) grows with time. If the largest scales can no longer grow (due to the simulation box size, or tunnel cross section) then a decay exponent of *n=2* is expected (Stalp *et al.* 1999, Touil *et al.* 2002). Some turbulence generation methods like the impinging jets used in Hwang and Eaton (2004) appear to generate very large scale turbulence and *n=2* decay right from the initial decay.

**1.2 Determining Decay Rates**

Some readers may be aware of the difficulties involved in accurately determining the decay rate for isotropic turbulence, but most readers will unaware that most experiments and simulations are not able to accurately determine the decay rate. To clearly describe the problems involved we briefly revisit the results of Batchelor and Townsend's classic experiment which are reproduced in Figure 1a. We choose this particular work because it was very carefully executed and because it is highly cited. However, the issues being raised by this experiment are not unique and are shared by much of the subsequent literature on turbulence decay. Figure 1b shows the results from Bennett and Corrsin (1978), which is another frequently cited low Reynolds number decay experiment.

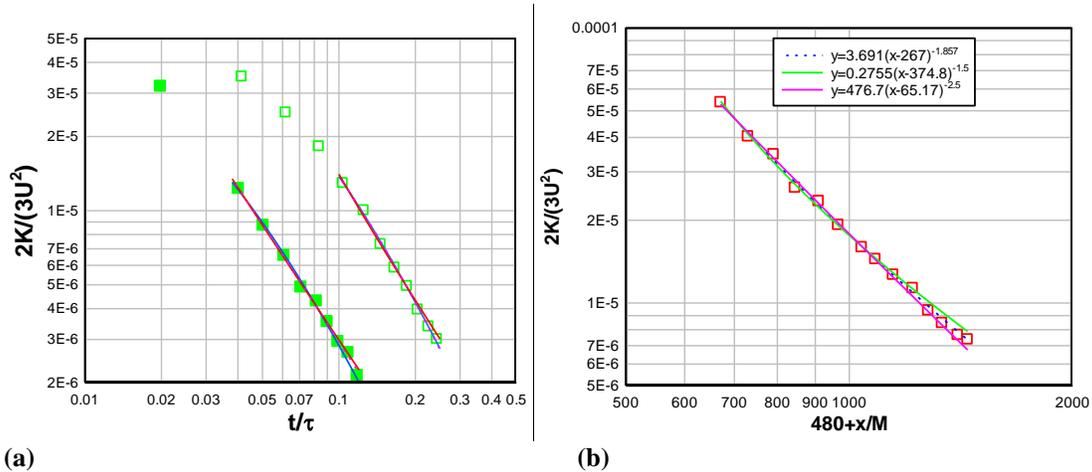

**(a)** **(b)**

**Figure 1.** Reproduced results from (a) Batchelor and Townsend and (b) Bennett and Corrsin for low Reynolds number isotropic decay. The data is insufficient to definitively decide between any of the proposed theoretical decay exponents.

On a log-log plot, power-law decay should appear as a straight line if the time origin is chosen correctly. Clearly, at very early times, the turbulence is settling down from its generation mechanism (which is actually wake turbulence) and the theory does not yet apply, but eventually a linear behavior appears. The slope of this line is the decay exponent. Batchelor and Townsend performed two experiments (doubling the wind tunnel speed from 6.2 m/s to 12.8 m/s). The lower left curve (solid boxes) is the higher speed (and higher Re) case and the low speed case is shown with open squares. We have chosen to assume that the first data point in the high speed case, and the first three data points in the low speed case are not yet on the line (i.e not in power-law decay). The choice of how to determine which points are not in the power law decay region is certainly an important issue, but it is not the crux of the issue we wish to raise. The real problem is that almost any decay exponent fits this data. We have plotted best fit curves to all three data sets (including Bennett and Corrsin). Our curve fits are not actually lines since the time offset is also an unknown in our fits. The lower line is the suggested value of *n=5/2*. It fits well. The upper line is *n=3/2* (which is a value proposed by a later theory developed by Saffman). It also fits the data reasonably well. Not too surprisingly, others have viewed this same data and found other values for the decay exponent. For example, Tan and Ling (1963) analyzed this same Batchelor and Townsend data set and concluded the exponent should be *n=2* for both speeds (which corresponds to their proposed theoretical analysis).

Letting the computer fit the exponent (as well as $\hat{K}$ and $t_0$) gives the dotted lines in between. The fact that it is difficult to tell any of these curves apart in the figure is exactly the difficulty we wish to emphasis in this section. The computer fit for the high speed case produces an exponent of 1.688, and the lower speed case produces an exponent of 2.675 and the Bennett and Corrsin experiment gives an exponent of 1.857 (using all the data in the fit). If we least-squares fit the low speed case with all the data points (including the first three) we find the exponent is 4.207, so maybe even the 4th and 5th data points should also be neglected (which then reduces the decay exponent to close to 1.5). Note that our computer aided curve fits are based on minimizing the least square error. The error is much larger where the turbulence is larger so these fits focus more on the early times and show the most differences at late times (where *K* and therefore its error are much smaller). Curve fits based on the relative error would give even different answers. Just as in the low Reynolds number case, most high Reynolds number experimental data sets support almost any theory reasonably well but can not differentiate between them. Skrbek *et al* (2000) review and replot a number of well known high Reynolds number data sets.

In summary, finding the decay exponent involves the *ad hoc* decisions of identifying what data is in power-law decay, and then how to weight that data (error, relative error, etc). Even more importantly, finding the decay exponent with a high enough degree of accuracy to make definitive statements about which theory is correct requires longer time measurements than these experiments provide and/or more accuracy in those measurements. Therefore, the precision of most existing decay exponent results (these and others) is too low to eliminate any of the existing theories as a possibility.

Note that it is very difficult to extend wind tunnel experiments to longer times. Besides requiring a very long test section, measurements at long times have four hurdles. The decay exponent is sensitive to: (1) when the turbulent lengthscale grows to the same order of magnitude as the wind tunnel cross section size, (2) when boundary layer turbulence and pressure fluctuations diffuse into the tunnel core, (3) when boundary layer growth is sufficient to cause effective axisymmetric contraction in the wind tunnel core and (4) when secondary mean recirculation zones in the duct corners (a trait of turbulent duct flow) becomes comparable to the turbulence levels (which decay to small values at long times).

It is perhaps one thing to question the experimental conclusions, but how can Batchelor and Townsend's exact solution to the statistical Navier-Stokes equations (the Karman-Howarth two-point correlation evolution equation) be in doubt? Agostini and Bass (1950) showed that there are at least an infinite set of solutions to the linearized (low Reynolds number) Karman-Howarth equation - of which the Batchelor and Townsend solution with *n=5/2* is one possibility and the Saffman low Re number prediction with *n=3/2* is yet another. Perhaps even more importantly, the whole idea of even linearizing the Karman-Howarth equation (neglecting the nonlinear terms entirely) might be called into question. For example, the limit of the Navier-Stokes equations at small viscosity (high Reynolds number) is not the same as finding solutions to the Navier-Stokes equations at zero viscosity (the Euler equations). The former produces boundary layers and the latter does not. Similarly, solutions of the two-point correlation equations (for turbulence) in the limit of small nonlinearity (low Reynolds numbers and small modal interaction) may be quite different from solutions with exactly zero nonlinearity (where modes do not interact at all). Perhaps we should not be curious about which of the Agostini and Bass solutions occurs in low Re number stirred turbulence, but even if any of them occur at all. For example, it appears that in the domain constrained case none of the Agostini and Bass low Reynolds number solutions are applicable.

**1.3 Relation of Decay Rates to Spectral Properties and Eddy Structure**

The various theories can be viewed in wavenumber space as differing in how the low wavenumber portion of the 3D energy spectrum behaves. Small wavenumber behavior that asymptotically behaves like $k^2$ for small k, produces the Saffman decay result and small wavenumber behavior of $k^4$ produces the Kolmorgorov result. The wavenumber analysis does not apply to bounded domains. Note that experimentally measuring the small wavenumber asymptotic behavior of the energy spectrum with sufficient accuracy is no easier than measuring the decay rates accurately.

The wavenumber analysis explains why classic numerical simulations of decaying turbulence are not helpful for determining decay rates of real fluids. In classic numerical simulations, the initial spectrum for the turbulence is imposed, or the turbulence is forced at low wavenumbers, or both. In either case, the low wavenumber behavior of the initial conditions or of the forcing dictates the low wavenumber spectrum and therefore the subsequent decay rates. If $k^2$ behavior exists in the spectrum, it can be shown that it will remain for all time. The Navier-Stokes equations (in the absence of boundaries) can not change that portion of the spectrum. The nonlinear terms in the Navier-Stokes equations can produce a low wavenumber $k^4$ contribution, so if the spectrum is initially $k^4$ or higher it will remain (or become) $k^4$ after some time. Therefore, in classical numerical simulations of isotropic turbulence the decay rate is invariably dictated by user choices not by the Navier-Stokes equations.

Davidson (2004) suggests that turbulence that consists of a collection of vortex rings (or eddies with no linear impulse) would have a $k^4$ spectrum and turbulence that consists of eddies with some linear impulse would have a $k^2$ spectrum. While this is an interesting framework for the problem this picture still does not make it clear what wind tunnel or stirred turbulence behaves like since we do not know which of these constituent units are present in stirred turbulence. Mathematically, it is not even clear how to decompose turbulence into these different types of eddies so this question can be answered.

One straightforward method to answer the question of which low-wavenumber spectrum or eddy type is present in stirred turbulence is to perform the simulations and observe the decay results. This is the approach taken in this paper.

**2. Simulation Methodology**

**2.1 Numerical Method**

Fourier spectral methods are very common for simulations of isotropic decaying turbulence. They have high order (exponentially convergent) spatial accuracy and are very fast because they can easily solve for the pressure and account for the incompressibility constraint (using discrete fast Fourier Transforms). We note however, that order of accuracy does not directly equate to actual accuracy when the mesh size is not small. In direct numerical simulations the mesh is as large as is possible such that the smallest physics is resolved. It is not small in the Taylor series approximation sense necessary to determine accuracy from mathematical arguments about the convergence order. In addition, Fourier Spectral simulations are limited in the boundary conditions that can be applied.

Since we now understand that the whole key to isotropic decay is actually the turbulence generation and whether a $k^2$ or $k^4$ low wavenumber spectrum forms, our choice of simulation technique revolves as much around the simulation of the turbulence generation as it does around the turbulence decay. Our particular interest in this work is in turbulence that has been generated by mechanical stirring. Since the generation mechanism is due to walls, a numerical method (staggered mesh discretization) was chosen that is known to very accurately capture the presence of walls (Perot and Moin, 1995). Cartesian staggered mesh methods are now fairly common in DNS simulations involving walls and geometric complexity. Cartesian staggered mesh schemes not only conserve mass and momentum to machine precision, but because they are a type of Discrete Calculus method (Perot & Subramanian, 2007) so they also conserve vorticity (or circulation) and kinetic energy in the absence of viscosity. As a result, there is no artificial viscosity /diffusion in this method except that induced by the time-stepping scheme (Perot 2000). Kinetic energy and vorticity conservation are important criteria in turbulence simulations if the energy cascade is to be captured correctly. In addition, the staggered mesh discretization is free from pressure modes and the need for pressure stabilization terms. Pressure from the walls is probably the reason that a particular large scale turbulence structure appears from relatively small scale mixing boxes, so it is important to compute this variable with high physical fidelity. These methods also treat the wall boundary condition well because the wall normal velocity unknown lies exactly on the wall, so no interpolation is required to enforce the kinematic no penetration condition. Higher order versions of this method exist (Subramanian & Perot, 2006), but have issues with boundary condition implementation and with parallelization. They did not produce superior results relative to their computational cost.

The solution method uses a three step, low storage Runge-Kutta scheme for time advancement that is second order accurate in time (Perot & Gadebusch 2009). This scheme is stable for eigenvalues on the imaginary axis less than 2, which implies CFL < 2 for advective stability. Our simulations always use a maximum CFL < 1. The diffusive terms are advanced with the trapezoidal method for each Runge-Kutta substep, and the pressure is solved using a classical fully discrete fractional step method (Perot, 1993), though an exact fractional step method (Chang *et al*. 2002) is also possible.

The simulations were performed on $512^3$ meshes (with roughly half a billion unknowns) with fully periodic boundary conditions. In general the domain is very large compared to other decay simulations at comparable Reynolds numbers. A very large domain in necessary in order for the turbulent length scale to be able to grow for long times before it reaches the domain size. The large domain size also allows the low wavenumber (large length scale) portion of the energy spectrum to be extremely well resolved compared to other DNS simulations. The large domain size limits the Reynolds numbers that can be obtained. Nonetheless, the highest Reynolds numbers simulated in this work are comparable to the Reynolds numbers found in laboratory wind tunnel experiments (such as Comte-Bellot and Corrsin 1971). In addition, the highest Reynolds numbers tested in this work are sufficient to show decay rates that are very consistent with high Reynolds number decay theories.

The computer does not require physical units, but physical units are helpful for the reader to put the simulations in perspective. If the simulated fluid is water at standard temperature and pressure (with $\nu = 10^{-6}$) then the domain size is a cube that is 48 cm on a side. The small cubes that stir the turbulence are 1.4 cm on a side. And in the $512^3$ simulations there are 768 of them randomly placed in the domain. The total volume of all the stirring elements is therefore 1.93% of the total simulation volume. The mesh size itself is 0.9375 mm (which is 1/15th of the stirring cube size). At early times in the simulation, the timestep can be as small as 1/1000th of a second. In all the simulations it is never larger than a 1/10th of a second. The simulations themselves run out to more than a thousand seconds (over 20 physical minutes). For comparison, residence times of more than 2 seconds are rare in wind tunnel experiments. During this time the turbulence level (as measured by the kinetic energy, *K*) drops roughly three orders of magnitude.

**2.2 Exponent Calculation**

In order the measure the decay exponent a methodology that requires no user input is sought. We do not wish to estimate $t_0$, decide which points are in power-law decay, decide how to weight the curve fits, or introduce other possibilities for uncertainty. In this work, the decay exponent is therefore calculated for the small time interval between any two measurement locations. Specifically, if it is assumed that a power-law decay exists in some small time interval, then in that interval $K(t) = \hat{K}(t+t_0)^{-n}$ and $\varepsilon(t) = -\frac{dK}{dt}(t) = n\hat{K}(t+t_0)^{-n-1}$ so $K/\varepsilon = \frac{1}{n}(t+t_0)$. The decay exponent in that small interval is therefore the inverse of the $K/\varepsilon$ slope.

In this work $K/\varepsilon$ is measured (from the DNS) at a large number of points in time. The analysis then assumes that between any two data points, the values of $\hat{K}$, $t_0$, and *n* are constant and therefore $K/\varepsilon$ is a linear function in that interval. The decay exponent is therefore determined for each interval between the data points. The value of n (and $\hat{K}$, and $t_0$) can of course change from interval to interval (and it does). But as the intervals become very small, the assumption of constant $\hat{K}$, $t_0$, and *n* in each interval becomes an increasingly good approximation. An example of what $K/\varepsilon$ looks like (for Run 6) is shown in Figure 2a. The corresponding inverse of the slope (*n*) is given on a log scale in time in Figure 2b. A log scale for time is appropriate for isotropic decay because as Figure 2a shows, the timescale for the flow increases significantly as time progresses and is roughly proportional to the time elapsed. There are almost 25,000 data points in each of these graphs, so the intervals are sufficiently small for the exponent calculations to be quite accurate.

The proposed method for determining the decay exponent is not as effective when considering experimental data, since it determines *n* from the derivative of a derivative ($\varepsilon$). Noise in the data can therefore be severely amplified by this double differentiation. This approach to the computation of the decay exponent is similar to that used by Burattini *et al.* (2006) though their approach is framed in terms of the Taylor microscale.

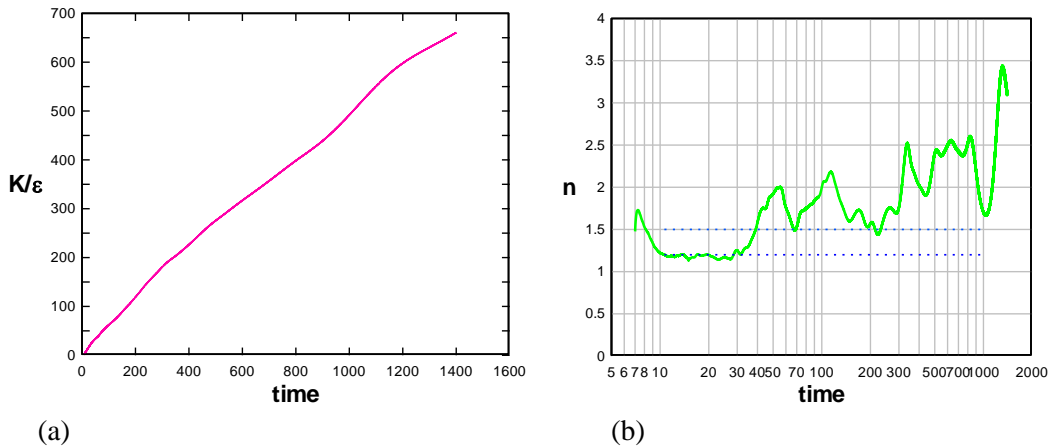

(a)     (b)

**Figure 2.** The turbulent timescale, $K/\varepsilon$, for Run 6 (which starts at a turbulent Re = 995) and the decay exponent (*n*) on a log scale. The horizontal dashed lines with the decay exponent are the (Saffman) high Re value of 6/5 and the low Reynolds number limit of 3/2.

In this work the value of the dissipation, $\varepsilon = \nu \int u_{i,j} u_{i,j} d\mathbf{V}$, is calculated from the DNS data at each time via a summation over the domain directly from the DNS velocity fields (the integral involves all 9 gradient terms). No assumptions about isotropy are used. Similarly, the kinetic energy, $K = \frac{1}{2}\int u_i u_i d\mathbf{V}$ is also calculated directly from the DNS data. The other option is to calculate $\hat{\varepsilon}$ by numerically differentiating the time series for $K$. Figure 3a shows the DNS data for $K$ and $\varepsilon$ (from Run 6). The values for $\hat{\varepsilon}_{2nd}$, which are calculated using $2^{nd}$ order (in time) central differentiation of $K$, are visually identical to $\varepsilon$, as is $K/\hat{\varepsilon}_{2nd}$. However, figure 3b shows the values of $n$ calculated from this data using the primary dissipation and using the $2^{nd}$ order derivative of the kinetic energy. This figure makes it clear that the decay exponent is an extremely sensitive quantity. The noise in the numerical differentiation shows up increasingly as

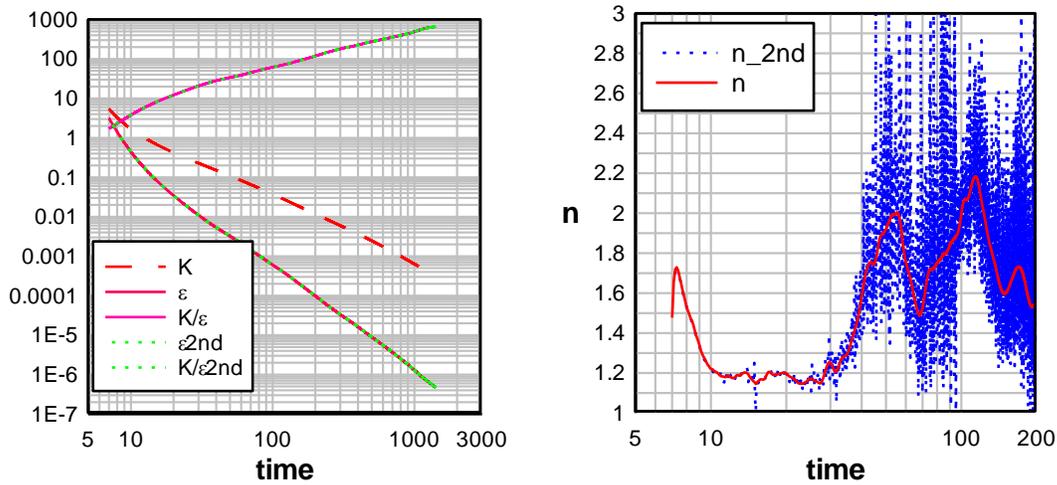

**Figure 3.** (a) Primary data, K, ε, and K/ε and derived dissipation $\hat{\varepsilon}_{2nd}$ for Run 6. (b) Decay exponent calculated from the dissipation and the numerically calculated dissipation.

K and $\varepsilon$ get smaller (and time proceeds). The noise in figure 3b at later times (t > 40) is similar to what happens even at early times when experimental data is doubly differentiated (using this method) to find $n$.

The sensitivity of the decay exponent is both a difficulty and a boon. Precise calculation of the decay exponent requires an extremely accurate simulation. But the decay exponent also provides an extremely sensitive test that a simulation (particularly the large scales) are actually being computed correctly. For example, the decay exponent is a far more sensitive quantity than the velocity derivate skewness which is $-0.5 \pm 0.016$ (3% maximum variation) for entire time (7 to 200) for Run 6. While skewness measures if the cascade is being well captured it apparently does not distinguish if the large scales (small wavenumbers) are being correctly represented.

**2.3 Turbulence Generation**

The generation of the turbulence is an important component of this work. For these simulations, 768 small no-slip cubes were randomly distributed throughout the domain. It is possible that two or more cubes can intersect with each other when they are randomly placed. This is allowed and is computed correctly, but it is rare because these small mixing cubes fill less than 2% of the total fluid volume. The cubes remain fixed and an external (constant in space) acceleration is applied to the fluid to drive it past the cubes. This is equivalent to performing the calculation in a linearly accelerating reference frame. The direction of this acceleration is random, but its magnitude is specified by the user. A typical value of the acceleration is 1 m/s$^2$ (or about 1/10 the acceleration of gravity). In these simulations the direction of the acceleration is changed to a new random direction (but of the same magnitude) every 0.3 seconds. This time scale is much less than the initial large eddy timescale (which is on the order of 2.0), but long enough to create a significant wake behind each cube.

This procedure produces a random walk in the acceleration or a 'shaking' of the fluid domain. However, a random walk does not remain bounded. In order to make the random accelerations more like shaking (with a mean acceleration of zero), another acceleration is imposed which is proportional to the current mean velocity $\mathbf{a}_{return}(t) = -\frac{\bar{\mathbf{u}}(t)}{0.3}$. If the primary acceleration stays too long in one direction this correction tends to become large enough to counteract it. The timescale is such that this term is large only if two or more random accelerations in a row happen to be in roughly the same direction.

The shaking is performed for 5.1 seconds in most simulations (or 17 different accelerations). The primary acceleration (shaking) is then turned off and the restoring acceleration $\mathbf{a}_{return}$ only is allowed to act. This is the final motion of the domain back to its rest position. After 1.9 seconds this restoring acceleration (which is exponentially decaying in time) causes the mean flow to be extremely close to zero. A mean flow of zero is not necessary for the code, but it does allow the simulation to take slightly larger timesteps (by minimizing the CFL stability criteria), and it does seem to lead to better statistical accuracy at very long times (when the fluctuations can become much smaller than the mean flow). During this 1.9 second period the turbulence changes from being accelerated to being in isotropic decay. At the end of this period (when the mean flow is zero), the boxes instantaneously turn into (zero velocity) fluid.

To get a sense of the size and density of the stirring boxes the zero streamwise velocity contour is shown in figure 4a. This shows the boxes, and some of the box wakes, in one 1/8$^{th}$ of the total simulation domain. The resulting turbulence, well after the boxes are gone, is shown in Figure 4b, where a slice through 1/8$^{th}$ of the domain is shown.

The decay officially starts at 7 seconds into the simulation. At early times after the boxes become fluid (at 7 seconds) there is an initial transient in the decay exponent that does not look like power law decay as the existing turbulence 'chews up' and mixes completely with the 2% of the domain where the mixing boxes used to be and stationary fluid was present. This takes about

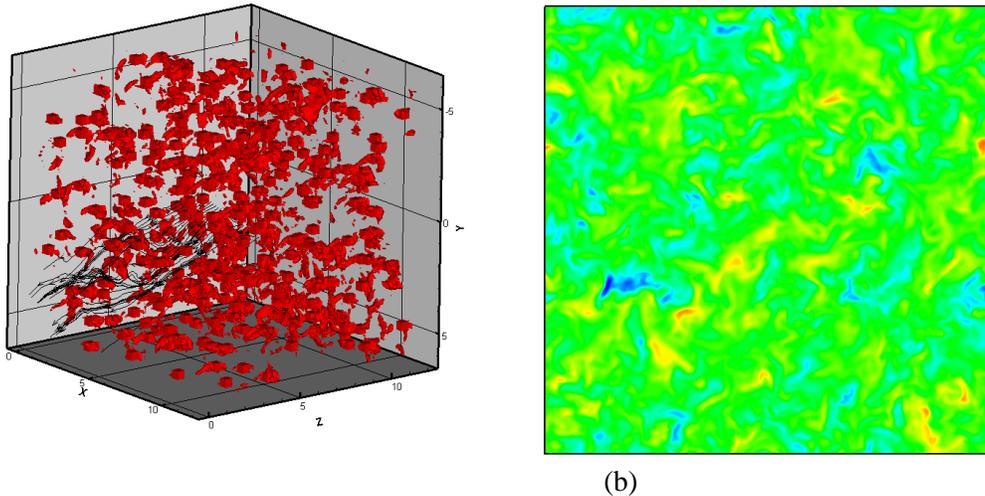

(a)                                                         (b)
**Figure 4.** (a) Zero velocity contour showing the mixing cubes (and some wakes) in 1/8$^{th}$ of the simulation domain. (b) Slice though the domain at time=15 (well after the turbulence is fully isotropic), showing the streamwise velocity. This is ¼ of the total calculation domain.

2.5 large eddy turnover times to complete (roughly 5 seconds on many of the simulations). It is not possible to leave the boxes as no-slip regions in the domain because these would significantly alter the decay process.

After 12 seconds the turbulence is generally in pure isotropic decay. The longitudinal and transverse structure functions are shown in Figure 5a. This data is from Run 6 (shown previously) at a time of 15 seconds. The lower solid line is the longitudinal structure function and the upper curve is the transverse structure function. The small dashed lines indicate various theoretical limits. At small separations the dotted line shows that the structure functions are proportional to $r^2$ and the smallest scales are well resolved (using three mesh units). At small r the transverse structure function is exactly twice the longitudinal one (as theory predicts). At about 20 mesh separations the structure functions have the inertial range behavior of $r^{2/3}$ and a difference in magnitude of 4/3 (as predicted by isotropic decay theory). The inertial range in this turbulence (at Re=559, $Re_\lambda$ = 61) is not large. But the experiments of Compte-Bellot and Corrsin (1971) are at nearly the same Reynolds number. Finally at large separations the structure functions both approach the constant theoretical value of $\frac{4}{3}K$. The large dashed curve is a prediction of the transverse structure function from the longitudinal structure function and theory. There is a small discrepancy on at very large separations. This is expected since the statistical sample decreases with separation distance. At r/Δx =128 (1/4 of the domain size) the data is being averaged over only roughly 64 very large eddies.

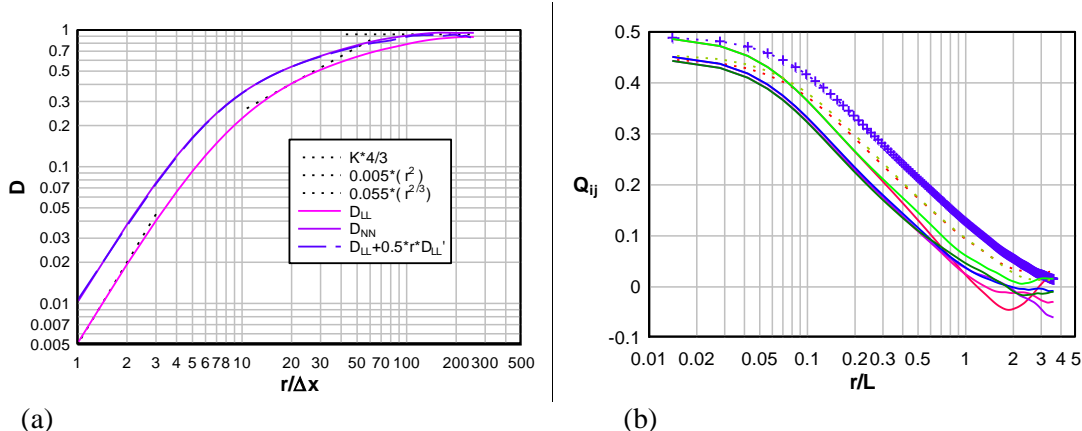

(a)          (b)

**Figure 5.** (a) Longitudinal and transverse structure functions for the turbulence in Run 6 at a time of 15. Dashed lines are various theoretical limits. (b) Two point correlations. Solid lines are transverse correlation and dashed lines are longitudinal correlations.

Figure 5b shows the corresponding two-point correlations. The dashed lines are the longitudinal correlations (3 total) and the solid lines are the transverse correlations (6 total). The correlations have not been normalized by their peak value so that the anisotropy in the turbulence can be seen. Due to statistical variability the turbulence is never perfectly isotropic in these simulations. The higher set of curves in this figure corresponds to the $Q_{22}$ correlations in this particular case. All the simulations showed similar small levels on anisotropy in the turbulence. The diagonal anisotropy ($a_{ij} = \frac{R_{ij}}{K} - \frac{2}{3}\delta_{ij}$) values for this case at this time are (-0.024, +0.037, -0.013). The off-diagonal values are slightly smaller. Similar anisotropy levels are found in wind tunnel experiments. These small levels of anisotropy do not appear to affect the kinetic energy decay rates. The large eddy length scale, $L = K^{3/2}/\varepsilon$ is 6.69 and the Taylor microscale, $\lambda_g = \left(\frac{10\nu K}{\varepsilon}\right)^{1/2}$ is 0.895 at this time. The inertial range (at roughly r/Δx = 20) is at r/L = 0.28.

There are, of course, a host of other statistics one might be interested in concerning this flow (spectra, flatness, and so on). The goal of this paper, however, is to discuss the turbulent decay rate. Since the decay rate is the most sensitive parameter governing this flow, and since the decay rate has firm theoretical limits, it is prudent to establish the decay rate first, before exploring more complex statistics of the turbulence in any more detail.

## 3. Results

Results from a total of 18 simulations are presented. Four of these simulations involve tests of the numerics, the other 14 involve variations in the initial condition and problem parameters. These 14 are tabulated below. Of these 14 cases, 10 involve $512^3$ meshes and 4 simulations were performed at lower Reynolds numbers and used $256^3$ meshes.

|  | $K_0$ | $\varepsilon_0$ | $Re_0$ | $Re_{\lambda,0}$ | $T_0 = \frac{K}{\varepsilon}$ | $L_\eta / \Delta x$ | $L_0 = \frac{K^{3/2}}{\varepsilon}$ | Stir_L | accel | mesh |
|---|---|---|---|---|---|---|---|---|---|---|
| Run 1 | 2.117 | 1.309 | 342 | 47.7 | 1.617 | 0.315 | 2.353 | 0.844 | 100 | $512^3$ |
| Run 2 | 3.868 | 2.249 | 665 | 66.6 | 1.720 | 0.275 | 3.384 | 1.406 | 100 | $512^3$ |
| Run 3 | 3.940 | 1.970 | 788 | 72.5 | 2.001 | 0.285 | 3.972 | 1.969 | 100 | $512^3$ |
| Run 5 | 2.961 | 1.550 | 566 | 61.4 | 1.911 | 0.302 | 3.287 | 1.406 | 75 | $512^3$ |
| Run 6 | 5.464 | 3.199 | 933 | 78.9 | 1.708 | 0.252 | 3.992 | 1.406 | 130 | $512^3$ |
| Run 7 | 0.508 | 0.209 | 124 | 28.8 | 2.432 | 0.499 | 1.734 | 1.406 | 40 | $512^3$ |
| Run 9 | 1.424 | 0.633 | 321 | 46.3 | 2.251 | 0.378 | 2.686 | 1.406 | 60 | $512^3$ |
| Run 10 | 2.269 | 1.209 | 426 | 53.3 | 1.879 | 0.322 | 2.829 | 1.406 | 60 | $512^3$ |
| Run 11 | 1.581 | 0.766 | 326 | 46.6 | 2.063 | 0.361 | 2.594 | 1.406 | 50 | $512^3$ |
| Run 14 | 0.270 | 0.101 | 72 | 21.9 | 2.668 | 0.598 | 1.387 | 1.406 | 30 | $512^3$ |
| Run J | 0.191 | 0.0846 | 43 | 16.9 | 2.256 | 0.625 | 0.986 | 0.844 | 20 | $256^3$ |
| Run K | 0.688 | 0.401 | 118 | 28.0 | 1.716 | 0.424 | 1.423 | 0.844 | 50 | $256^3$ |
| Run L | 1.912 | 1.205 | 303 | 44.9 | 1.586 | 0.322 | 2.193 | 0.844 | 100 | $256^3$ |
| Run M | 1.392 | 0.803 | 241 | 40.1 | 1.735 | 0.356 | 2.047 | 0.844 | 100 | $256^3$ |

**Table 1.** Summary of the initial conditions (at the time when the stirring boxes disappear) for the different simulations presented in this paper.

The first three runs look at the affect of the variation in the box size while the total volume of the stirring boxes remains constant at just under 2%. The next two cases (and Run 2) examine the behavior at high Reynolds numbers. Run 7, Run 14 (and Run1) look at low Reynolds number behavior, and the four $256^3$ cases confirm those low Re results. The final three $512^3$ simulations (Runs 9, 10 and 11) look at transition from high to low Reynolds number behavior. The smaller simulations also serve as a mesh resolution study since Run L and M are essentially the same as Run 1, but Run 1 has twice the resolution. The Reynolds number in Table 1 and throughout the text is the turbulent Reynolds number $Re = \frac{K^2}{\nu\varepsilon}$ since this is what is more commonly used in turbulence models. In isotropic turbulence the Taylor microscale Reynolds number is related to the turbulent Reynolds number by $Re_\lambda = (\frac{20}{3} Re)^{1/2}$, which is provided in the table because this metric is commonly used in isotropic decaying turbulence simulations. The Kolmogorov length scale $L_\eta$ starts at roughly 1/3 of the mesh size, but always is ½ or greater (even for Run 6) by the time the turbulence is in power law decay (after roughly 2 large eddy turnover times). A value greater than ½ is considered to be more than sufficient small scale resolution by most DNS practitioners.

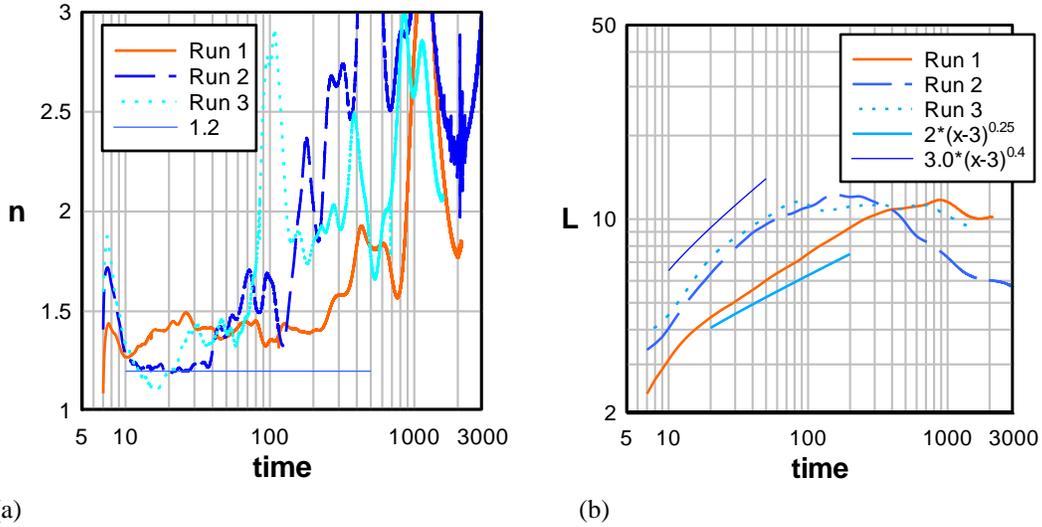

**Figure 6:** (a) decay exponent and (b) large-eddy length scale $L = \frac{K^{3/2}}{\varepsilon}$ versus time for different initial stirring box sizes.

### 3.1 Effect of Stirring Box Size

The affect of the stirring box size on the decay exponent is shown in figure 6a. Run 1 has 3584 boxes that are $9^3$ grid cells in dimension (1.95% of the domain volume). Run 2 has 768 boxes that are $15^3$ grid cells in dimension (1.93% of the domain volume). And Run 3 has 256 stirring boxes that are $21^3$ grid cells in dimension (1.77% of the domain volume). All these simulations were run in parallel on 128 cores. For load balancing reasons every core has the same number of stirring boxes. For Run 3 this means each core computes the effect of only two stirring boxes, so while the box placement is random within the CPU subdomain, the stirring boxes themselves are fairly evenly distributed throughout the whole domain. For Run 1, each core has 28 stirring boxes, which may lead to less uniformity in the box distribution statistics.

The decay exponent results look erratic at first glance, but begins to make sense after close inspection. Up until about a time of 13 (roughly three large eddy turnover times, $T = \frac{K}{\varepsilon}$, after the decay starts) there is a development phase. This is where the stationary lumps of fluid, where the stirring boxes used to be, get stirred into the rest of the turbulent flow. In these three cases, the development causes an overshoot in the decay exponent, but an overshoot does not always occur. The overshoot in all three of these runs is probably due to the fact that these runs have very similar (though not identical) shaking accelerations applied to them. After the development phase, Run 1 slowly approaches the (Saffman) low Reynolds number value of 3/2. It remains near that value for many turnover times until it increases at time 250 towards higher values. Run 2 obtains very close to the (Saffman) high Reynolds number value of 6/5 for some time. At time 40 it changes to the low Re value and at time 150 it heads higher. Finally, run 3 spends less time around a value of 6/5. It then moves towards 3/2 (the Re value) at t=25 and even higher at t=70. The Batchelor /Kolmogorov theory predicts a decay exponent of 1.428 at high Re numbers rising up to 2.5 at low Re numbers.

The departure of n from a single constant value at long times can best be explained by the large eddy lengthscale, $L = \frac{K^{3/2}}{\varepsilon}$, which is shown in Figure 6b for the same runs. On a log-log scale (with the time axis shifted appropriately to $t+t_0$) the lenthscale should grow linearly. Under the Saffman theory the lengthscale should have a slope of 2/5 at high Re and ¼ at low Re. Under the Kolmorgorov and Batchelor theories it should be 2/7 at high Re and ¼ at low Re (meaning almost no change with Re). The figure shows lengthscale growth at short times that is roughly equivalent to the Saffman prediction and then a capping of the large eddy lengthscale at longer times. The maximum the turbulent lengthscale can achieve appears to be roughly ¼ of the domain size. The lengthscale maximum is reached at roughly t=250 for Run 1, t=150 for Run 2 and t=80 for Run 3. This corresponds very closely to the times when the decay exponent becomes very noisy and moves roughly towards a value of 2. Remember, n = 2 is the theoretical value expected from high Reynolds number size-constrained turbulence (see Appendix A).

These results indicate that the largest stirring box size (21 mesh units, or almost 2cm) is likely to have a large eddy lengthscale that approaches the domain limit size (12 cm) fairly quickly. This limits the effectiveness of simulations with stirring boxes this large. The smaller stirring box sizes, however, are adequate, and allow reasonably long times to be calculated before the turbulence lengthscales are constrained by the domain size.

### 3.2 Improving Long Time Results

The decay exponent is a very sensitive quantity, so perhaps the noise at long times is statistical variation or due to numerical error. In order to confirm this we tested a number of numerical parameters. Run 4 used RK4 instead of RK3 (no appreciable difference). Run 8 used 10 times smaller tolerances on the Conjugate Gradient solver (no significant difference). Run 13 made sure the pressure had a zero mean value at all times (somewhat smoother). Run 12 ran the decay process for 2.0 seconds longer so that the mean velocities (which were on the order of 0.02 cm/s in Run 2) were 35 times smaller than in Run 2. These last two changes are shown in Figure 7. They improve the long time behavior significantly, and make it clearer that the value that the exponent is approaching is probably 2 when the numerics are better. The lengthscale behavior is also improved. Run 12 has a constant lengthscale after it reaches the maximum value of ¼ of the domain size. This result indicates that the long time behavior (t > 300) of Runs 1-11 may also be noisier than is necessary at long times due to the numerics.

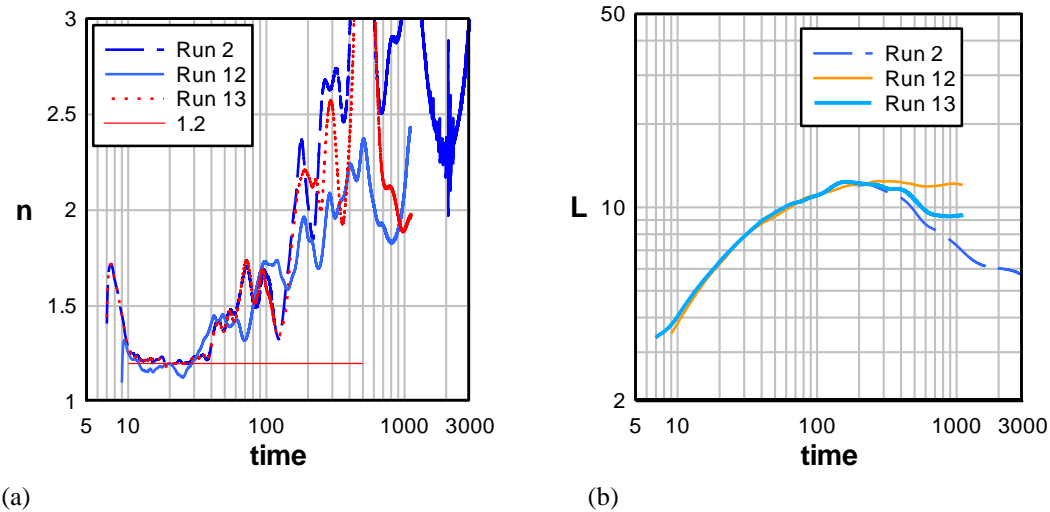

**Figure 7:** (a) decay exponent and (b) large-eddy length scale $L = \frac{K^{3/2}}{\varepsilon}$ versus time. All the conditions are the same as Run 2. Run 12 decays for an extra 2.0 seconds (and starts with a smaller mean flow). Run 13 keeps the mean pressure at 0 for all time.

### 3.3 High Reynolds Number

The primary affect of the changing the acceleration magnitude during the stirring phase is to change the initial Reynolds number of the turbulence. The decay exponent for three different accelerations (75, 100, and 130 cm/s$^2$) is shown in Figure 8a. The turbulent Reynolds number ($Re = \frac{K^2}{\nu\varepsilon}$), and the turbulent lengthscale ($L = \frac{K^{3/2}}{\varepsilon}$), are shown in figure 8b. Runs 5 and 6 have the same stirring box placement and nearly the same acceleration directions. Run 12 (which is the better version of Run 2) has a different initial box placement and acceleration directions, which just happen to give it properties very similar to Run 5 (even though the acceleration magnitude is different). All three simulations use the intermediate stirring box size of 15 grid cells (1.406 cm). These runs all show essentially high Reynolds number decay, until the lengthscale becomes domain size constrained (at t=30 for Run 6, and t=70 for Run 5 and Run 12). The slightly lower Reynolds number cases (Run 12 and Run 5) appear to begin transitioning to low Re decay behavior at Re = 250 (at about t=40). This transition to low Re decay is not complete when those simulations becomes domain size constrained at t=70.

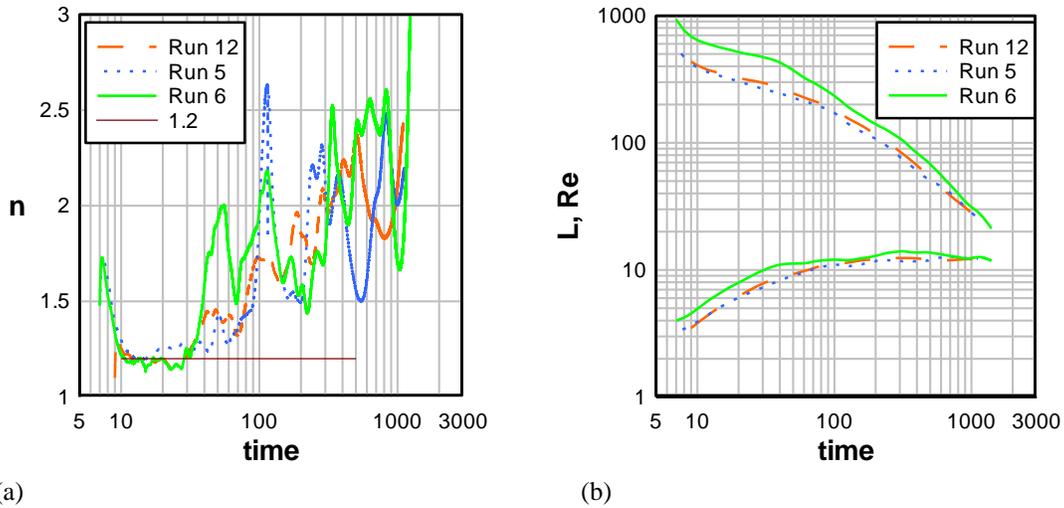

(a)            (b)

**Figure 8:** (a) decay exponent and (b) large-eddy length scale, Reynolds number $Re = \frac{K^2}{\nu\varepsilon}$ and negative skewness, versus time for the higher Reynolds number cases.

The velocity derivative skewness was computed for Run 12. It starts (at t=9) at a value of -.489 and increases monotonically to a value of -.512 (at t=150). This is very close to the accepted value of -0.5 for this Reynolds number (Re = 400, $Re_\lambda$ = 52) from Sreenivasan and Antonia (1997). These results for the skewness indicate that it is not a particularly useful quantity for determining if the turbulence is in power law decay. The skewness does not indicate when the domain size constraint changes the nature of the turbulent decay.

**3.4 Low Reynolds Number**

Three lower Reynolds number cases (Run 1 from before, along with Run 7, and Run 14) are plotted in Figure 9. Run 7 and Run 14 have identical stirring box locations and acceleration directions. Only the magnitude of the acceleration differs (from 40 to 30 cm/s$^2$). The initial Reynolds numbers are quite different for these runs (342, 130, and 72) but the decay exponent is very similar for all three cases and hovers around the value of 1.4. In all three cases the Reynolds number is below 200 by the time the initial transient is complete. In all three cases the decay exponent consistently stays just below the theoretical low Reynolds number value of 1.5 at early times (t < 100). For Run 1 the movement to n = 2 is clearly seen to be a result of the large eddy lengthscale becoming bounded (with L=10 at t=200). However, the two lower Reynolds number cases have large-eddy lengths that do not appear to be domain bounded (up to t < 1000), but their decay rates are rising above 1.5 at later times (t > 200).

It is possible that at low Re (Re < 50) the large eddy lengthscale is not the important physical length scale. Some viscous length scale may become important for determining when the turbulence is domain size constrained. It is also possible, that the low Re decay exponents are not approaching n = 2, but are instead beginning the transition to exponentially dominated decay ($n \rightarrow \infty$), which is another possible low Re limit. It is also possible that the simulations are too statistically sensitive at these later times to be definitive.

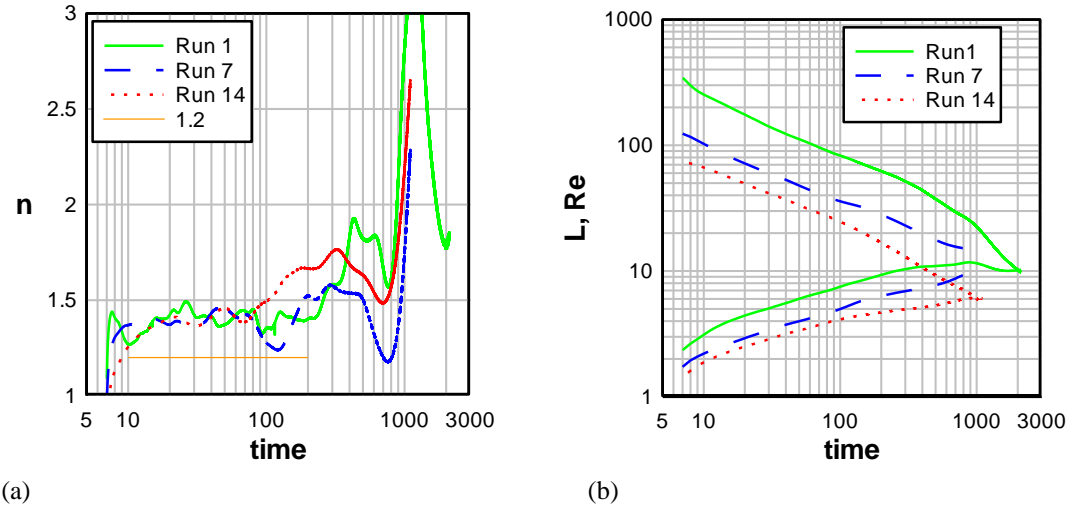

**Figure 9:** (a) decay exponent and (b) large-eddy length scale and Reynolds number $\text{Re} = \frac{K^2}{\nu\varepsilon}$ versus time for the low Reynolds number cases.

Four graphics processor (GPU) accelerated $256^3$ simulations were performed (Menon and Perot 2007) to see if the decay exponent consistently resides below 3/2 or if this is a statistical anomaly due to the box placement or that particular acceleration schedule. These smaller mesh simulations have a domain which is 1/8 the volume of Run 1, and the number of stirring boxes (448) is 1/8 that found in Run 1 (3584). They are therefore practically the same simulation, with 1/8 the statistical sample size, and therefore a shorter time until the flow becomes domain size constrained. The results for these simulations are shown in Figure 10.

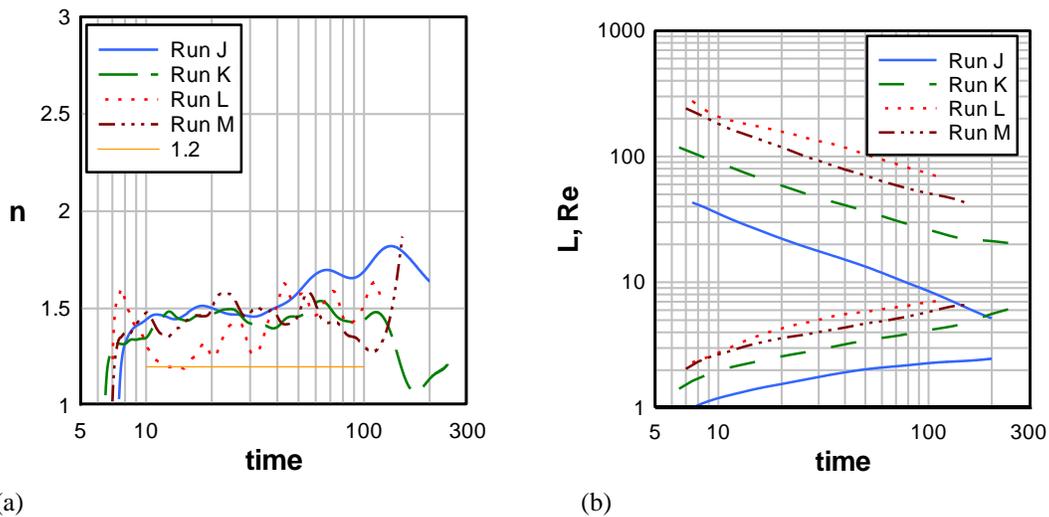

**Figure 10:** (a) decay exponent and (b) large-eddy length scale and Reynolds number $\text{Re} = \frac{K^2}{\nu\varepsilon}$ versus time for the low Reynolds number $256^3$ cases.

The domain-size limit on the large-eddy lengthscale should now be ½ what it was for the larger simulations (or roughly 6 cm). The highest Reynolds number case (Run L) shows the lengthscale flattening slightly around this value (t = 40). This case also shows high Re decay exponent behavior at early times (for Re values around 200) though it is also possible this is undershoot from the initial settling process. The other cases show the low Reynolds number decay exponent value to be very close to 3/2 for all these cases. The lowest Re number case (Run J) is very smooth and close to 3/2. The lengthscale appears to be flattening around t=40, but at a value of the lengthscale less than we saw for the high Re domain constrained cases. This is likely a Re affect as the Re (around 10 at this point in time) is now very low. We were not able to determine a criteria that consistently describes when the low Re simulations become box constrained

**3.5 Intermediate Reynolds Number**

The last three test cases, (Runs 9, 10 and 11) shown in Figure 11, were performed to try and isolate the transition process between the high Re number and the low Re decay states. This transition process has been studied numerically (Huang and Leonard 1994, Mansour & Wray 1994, Chasnov 1997) and experimentally (Kang *et al.* 2003, Lavoie *at al.* 2005). The Reynolds number decays with time so this transition event must occur eventually in any decaying turbulence. Run 10 and Run 11 have identical stirring box placement and acceleration directions. They vary only slightly in their Reynolds number. Run 9 has a similar Re number to Run 10 but uses different initial conditions.

These cases show the domain-size constraint occurring at the same time (t > 200). For each case the decay exponent starts close to the high Re value (6/5) and moves to the low Re value (3/2) well before the domain constraint becomes effective. Run 10 which is a slightly higher Re version of Run 11 (but otherwise identical) transitions slightly later than Run 11, which is what might be expected. However, Run 9 which has the same Re as Run 10 but different initial conditions transitions much earlier. It appears the transition process from high to low Re is Reynolds number dependent but also highly statistically variable. A similar situation that shows this kind of statistical variation superimposed on a Reynolds number dependence is seen in the relaminarization of pipe flow.

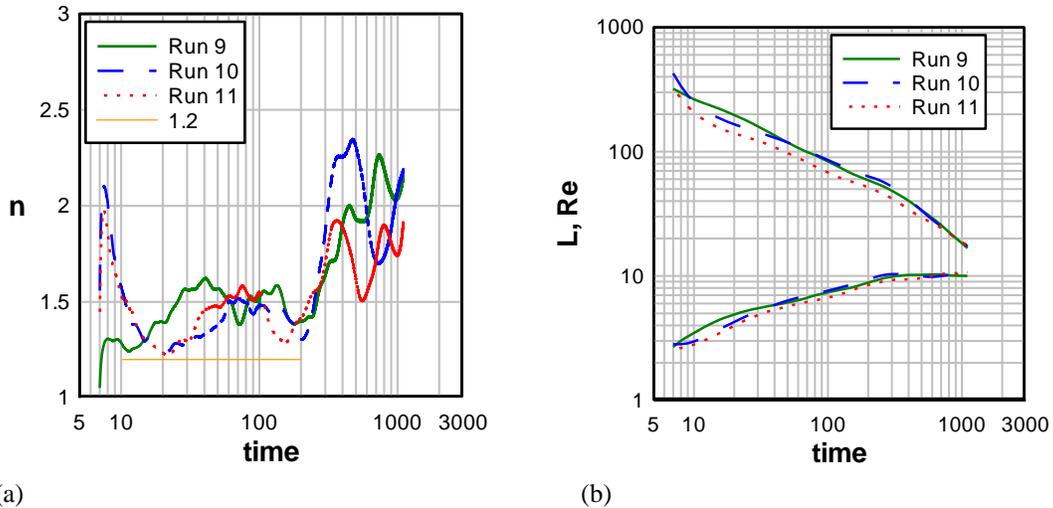

(a)  (b)

**Figure 11:** (a) decay exponent and (b) large-eddy length scale and Reynolds number $\mathrm{Re} = \frac{K^2}{\nu\varepsilon}$ versus time for the intermediate Reynolds number cases.

The change in the Reynolds number as the decay exponent goes from its high Re to its low Re values is relatively small (on the order of 30%). This suggests, that the decay exponent transition process seen in Figure (9a) is not being dictated by the change in the Reynolds number. Rather, the transition is essentially abrupt, but takes some time to fully percolate through the energy cascade and manifest itself in the decay exponent. The time for the transition to occur appears to be roughly the same for both Run 9 and Run 11 on the log(time) scale. This suggests that the transition from high Re values to low Re values occurs on the scale of the large eddy turnover time. At t=15 (Run 9) two large eddy turnover times is 15 s, and at t=30 (Run 10 and 11) two eddy turnover times is 40 s. These times are very close to the decay exponent transition times observed in figure 11a. It is hypothesized that the transition from high Reynolds number decay to low Reynolds number decay becomes increasingly likely as the turbulent Reynolds number drops below 250 (or as the Taylor microscale Reynolds number drops below 41).

**4. Discussion**

The decay exponent for turbulence with zero mean momentum and zero mean vorticity has been accurately computed for a wide variety of conditions. Very long time simulations allow each simulation to cover a number of regimes. Low Reynolds number, high Reynolds number, and domain constrained regimes have been clearly identified. Domain constraint occurs in these simulations when the large eddy lengthscale (L=$K^{3/2}/\varepsilon$) approaches ¼ of the domain size. This is roughly equivalent to the longitudinal correlations scale ($L_{11} = \int_0^\infty f(r)dr$) on the order of 1/8$^{th}$ the domain size and the transverse correlations scale ($L_{22} = \int_0^\infty g(r)dr$) on the order of 1/16$^{th}$ of the domain size.

Perhaps the most striking result of this work is the consistency of the results at both high and low Re with the isotropic decay theory of Saffman. The Saffman theory is sometimes discounted as a possible solution because the Fourier transformed Navier-Stokes equations can not produce the $k^2$ low-wavenumber spectrum (which is necessary for Saffman type decay). This objection is invalid for these results because the Fourier transformed Navier-Stokes equations neglect the influence of boundaries, and boundaries are precisely how the turbulence in this work (and in all mechanically generated turbulence) is generated.

The goal of this work was to resolve the critical turbulence modeling issue concerning the theoretical decay exponent for mechanically generated isotropic turbulence. This work suggests that in the limit of high Reynolds number decaying isotropic turbulence, the K/ε model constant $C_{\varepsilon 2}$ should be 11/6 = 1.833, and the low Reynolds number value should be 3/2 = 1.5. It is important to note that these limits are proven only for turbulence stirred using our particular method, but that there is every reason to believe that similar results will be achieved by any other type of mechanical stirring where turbulence is generated by walls (and the resulting pressure/incompressibility constraints). However, for turbulence that is *not* mechanically stirred there is no reason to believe that the Saffman result must still hold true. It is possible that other turbulence generation mechanisms could produce turbulence that follows the Batchelor-Kolmorgorov theory.

One inference of this work is that walls induce a $k^2$ behavior for the low wavenumber portion of the turbulence spectrum. We suspect that this is due to the wall blacking effect and resulting pressure signals. Another inference is that walls produce eddies which have some linear impulse.

The simulation results suggest that the transition from high to low Re number decay occurs quite differently from how it has been modeled in the past. All previous modeling efforts to capture the change in the decay exponent with Reynolds number have assumed that the two mixing processes (viscous and turbulent) act together, though at differing magnitudes (given by the Re), causing a gradual change in the decay exponent as the Reynolds number changes (see Perot and de Bruyn Kops 2006 and references therein). This leads to models in which the decay exponent (or $C_{\varepsilon 2}$) is a smooth function of the Reynolds number. In contrast, these simulation results suggest that the dynamics (or strange attractors) associated with turbulent decay change abruptly (and stochastically) around Re=250 ($Re_\lambda = 41$). It then takes a few eddy turn-over times for the turbulence to fully adjust to the new dynamics and produce the new decay exponent. It is not impossible for RANS type turbulence models to exhibit this type of rapid transition in the flow dynamics (Wang and Perot, 2002) if the model equations are structured correctly.

It should also be noted that this abrupt change in the dynamics (and thus the decay exponent) is not unique to the high-to-low Re number transition. A similarly abrupt transition appears to occur when the turbulence becomes domain-size constrained. The length scale tends to grow very linearly on a log-log plot and then very quickly becomes constant. There is not a gradual merger from one state to the other. The adjustment time in the decay exponent to the constrained (n =2) value is again on the order of a few eddy turnover times (see Run 12, Fig. 5 and Fig. 6).

## Acknowledgements

The bulk of these calculations were performed using NSF Teragrid supercomputing resources (Ranger) provided by the Texas Advanced Computing Center at the University of Texas at Austin. This work was supported in part by the Office of Naval Research and the Dept. of Defense via a subcontract from the Oak Ridge National Laboratory.